\documentclass[a4paper,fleqn,usenatbib,useAMS]{mnras}

\usepackage{graphicx}	% Including figure files
\usepackage{amsmath}	% Advanced maths commands
\usepackage{amssymb}	% Extra maths symbols
\usepackage{multicol}        % Multi-column entries in tables
\usepackage{bm}		% Bold maths symbols, including upright Greek
\usepackage{pdflscape}	% Landscape pagess
\usepackage[encapsulated]{CJK}
\usepackage{ucs}
\usepackage[utf8x]{inputenc}
\usepackage{multirow}
\usepackage{booktabs,caption}
\usepackage[flushleft]{threeparttable}
\usepackage[dvipsnames]{xcolor}
\usepackage[shortlabels]{enumitem}

\usepackage[dvipsnames]{xcolor}
\usepackage{tikz,hyperref}

\definecolor{lime}{HTML}{A6CE39}
\DeclareRobustCommand{\orcidicon}{
	\begin{tikzpicture}
	\draw[lime, fill=lime] (0,0) 
	circle [radius=0.13] 
	node[white] {{\fontfamily{qag}\selectfont \tiny ID}};
	\draw[white, fill=white] (-0.1625,0.00) 
	circle [radius=0.007];
	\end{tikzpicture}
	\hspace{-2.5mm}
}

\foreach \x in {A, ..., Z}{\expandafter\xdef\csname orcid\x\endcsname{\noexpand\href{https://orcid.org/\csname orcidauthor\x\endcsname}
			{\noexpand\orcidicon}}
}
\usepackage[T1]{fontenc}
\usepackage{ae,aecompl}

\usepackage{newtxtext,newtxmath}
\title[Shaping the H-deficient ejecta of A30]{Common envelope evolution in born-again planetary nebulae - Shaping the H-deficient ejecta of A\,30}

\author[J.~B.\,Rodr\'{i}guez-Gonz\'{a}lez et al.]
{J.~B.\,Rodr\'{i}guez-Gonz\'{a}lez\thanks{E-mail: j.rodriguez@irya.unam.mx}$^{1\orcidA}$, E.~Santamar\'\i a$^{2,3\orcidB}$, J.~A.\,Toal\'{a}\thanks{Visiting astronomer at Instituto de Astrof\'{i}sica de Andaluc\'{i}a (IAA-CSIC, Spain) through the Centro de Excelencia Severo Ochoa Visiting-Incoming programme.}$^{1\orcidC}$, M.~A.\,Guerrero$^{4\orcidD}$, 
B.\,Montoro-Molina$^{4\orcidE}$, 
\newauthor{G.\,Rubio$^{2,3\orcidF}$, D.\,Tafoya$^{5\orcidG}$, Y.-H.\,Chu$^{6\orcidH}$, G.\,Ramos-Larios$^{2,3\orcidI}$ and L.\,Sabin$^{7\orcidJ}$}
\vspace{1.5mm}
\\
$^1$Instituto de Radioastronom\'{i}a y Astrof\'{i}sica, UNAM, Antigua Carretera a P\'{a}tzcuaro 8701, Ex-Hda. San Jos\'{e} de la Huerta, Morelia 58089, Mich., Mexico\\
$^2$CUCEI, Universidad de Guadalajara, Blvd. Marcelino Garc\'\i a Barrag\'an 1421, 44430, Guadalajara, Jalisco, Mexico \\
$^3$Instituto de Astronom\'\i a y Meteorolog\'\i a, Depto.\ de F\'\i sica, CUCEI, Av.\ Vallarta 2602, 44130, Guadalajara, Jalisco, Mexico\\
$^4$Instituto de Astrof\'\i sica de Andaluc\'\i a, IAA-CSIC, Glorieta de la Astronom\'\i a s/n, 18008, Granada, Spain\\
$^5$Department of Space, Earth, and Environment, Chalmers University of Technology, Onsala Space Observatory, SE-439 92 Onsala, Sweden\\
$^{6}$Institute of Astronomy and Astrophysics, Academia Sinica (ASIAA), No. 1, Section 4, Roosevelt Road, Taipei 10617, Taiwan\\
$^{7}$Instituto de Astronom\'{i}a, Universidad Nacional Autonoma de M\'{e}xico, Apdo. Postal 106, 22800 Ensenada, B.C., Mexico
}

\pubyear{2022}

\begin{document}
\label{firstpage}
\pagerange{\pageref{firstpage}--\pageref{lastpage}}
\maketitle

\begin{abstract}
Born-again planetary nebulae (PNe) are extremely rare cases in the evolution of solar-like stars. It is commonly accepted that their central stars (CSPN) experienced a {\it very late thermal pulse} (VLTP), ejecting H-deficient material inside the evolved H-rich PN. Given the short duration of this event and the fast subsequent evolution of the CSPN, details of the mass ejection are unknown. 
We present the first morpho-kinematic model of the H-deficient material surrounding a born-again PN, namely A\,30. New San Pedro M\'{a}rtir observations with the Manchester Echelle Spectrograph were recently obtained to map the inner region of A\,30 which are interpreted by means of the software {\sc shape} in conjunction with \emph{HST} WFC3 images. The {\sc shape} morpho-kinematic model that best reproduces the observations is composed by a disrupted disk tilted $37^\circ$ with respect to the line of sight and a pair of orthogonal opposite bipolar ejections. We confirm previous suggestions that the structures closer to the CSPN present the highest expansion velocities, that is, the disrupted disk expands faster than the farther bipolar features. We propose that the current physical structure and abundance discrepancy of the H-deficient clumps around the CSPN of A\,30 can be explained by a common envelope phase following the VLTP event. Our proposed scenario is also compared with other known born-again PNe (A\,58, A\,78, HuBi\,1 and the Sakurai's Object). 
\end{abstract}


%%%%%%%%%%%%%%%%%%%%%%%%%%%%%%%%%%%%%%%%%%%%%%%%%%

%%%%%%%%%%%%%%%%% BODY OF PAPER %%%%%%%%%%%%%%%%%%

\section{Introduction}
\label{sec:intro}

The planetary nebula (PN) Abell 30 (PN A66 30 or PN G\,208.5+33.2; hereinafter A\,30) is a H-deficient PN that is accepted to have experience a {\it born-again} event $\lesssim$900~yr ago \citep{Fang2014}. It exhibits a double-shell morphology that includes an outer H-rich almost spherical shell with a radius of $\sim$1 arcmin and an inner H-deficient cloverleaf-like structure that extends $\sim$30~arcsec from the central star \citep[CSPN;][]{Jacoby1979}.

\begin{figure}
\begin{center}
\includegraphics[width=\linewidth]{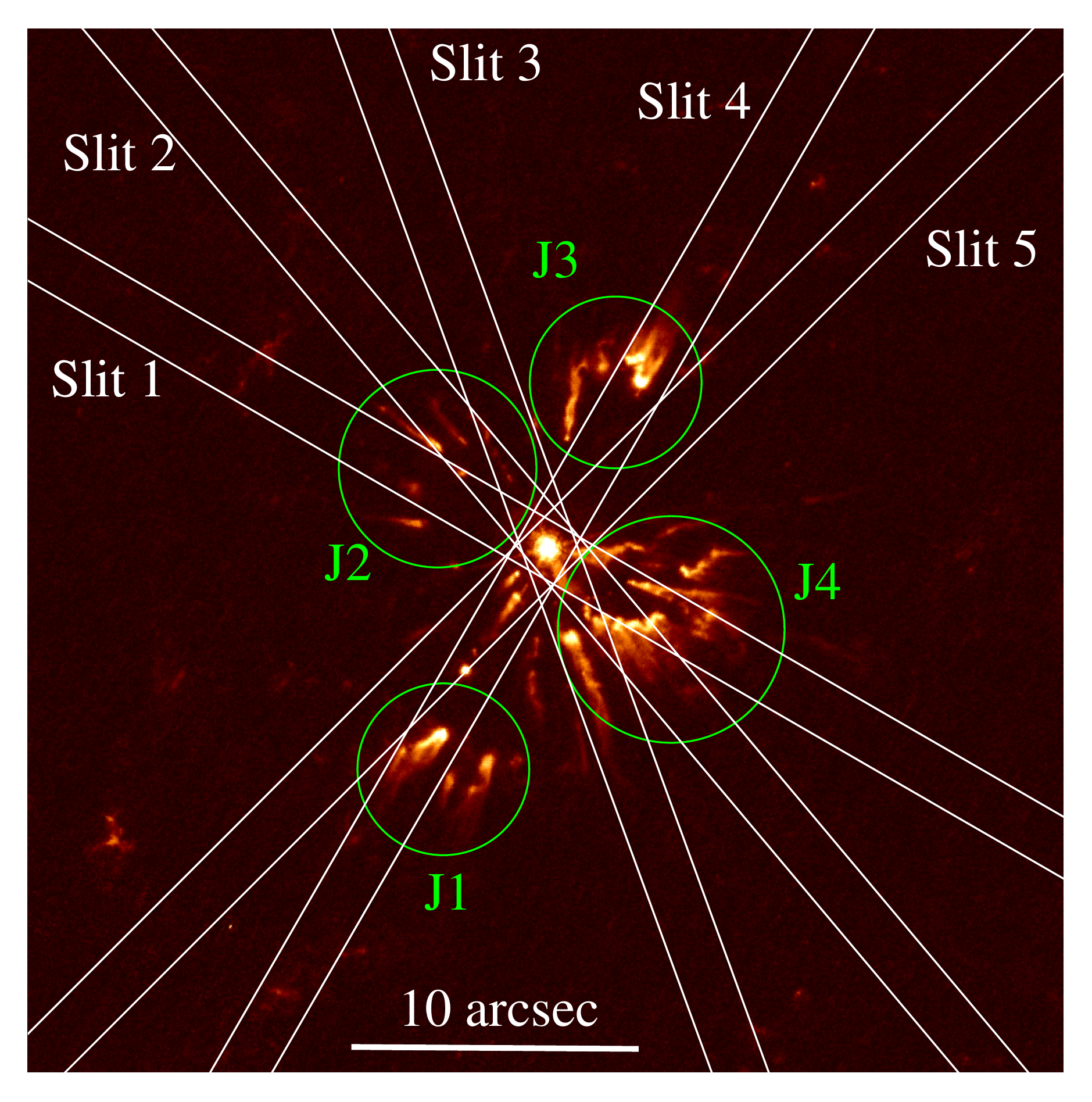}
%\vspace*{-0.4cm}
\caption{{\it HST} [O\,{\sc iii}] 5007~\AA\, narrow-band image of the central  region surrounding the CSPN of A\,30. The position of the SPM MES slits are illustrated and labelled as Slits 1--5 (see details in Table~\ref{tab:obs}). Green circles delineate the previously unresolved clumps defined as J1--J4 \citep{Reay1983}. North is up, east to the left.}
\label{fig:slits}
\end{center}
\end{figure}

The brightest H-deficient clumps are located $\lesssim$10~arcsec from the CSPN. Initial low-resolution optical images disclosed four main structures which were labelled as knot 1, 2, 3 and 4 \citep{Jacoby1979}, but these have been demonstrated later to be a collection of smaller clumps and filaments through high-spatial resolution {\it Hubble Space Telescope} ({\it HST}) images \citep[see][and Fig.~\ref{fig:slits}]{Borkowski1993,Borkowski1995}. 
In the following we will adopt the notation initially defined by \citet{Reay1983} of J1, J2, J3 and J4 for the knots, which correspond to the structures located towards the SE, NE, NW and SW from the CSPN of A\,30, respectively. These are highlighted with green circles in Fig.~\ref{fig:slits}.

There have been several spectroscopic studies of the H-deficient clumps in A\,30 which suggest that there are noticeable differences between knots J1--J3 and those of J2--J4 \citep[e.g.,][]{Jacoby1983,Guerrero1996}. The work of \citet{Simpson2022}, in combination with results presented by \citet{Wesson2003}, showed that there are also differences in the estimates of the abundance discrepancy factors (ADFs)\footnote{The ADF is a widely-known problem in ionized nebulae originally reported by \citet{Wyse1942} consisting in the discrepancy between the chemical abundances derived from recombination lines and collisionally excited lines. Typical values of the ADFs are around 2 \citep[see][]{Wesson2018}.} between knots. These works showed that the ADF for J4 is estimated to have an upper limit of 35, whilst that of J1 is estimated to be $\sim$700. However, it appears that without the information of the H mass fraction in the different clumps one cannot make strong statements on possible abundance differences. \citet{Simpson2022} concluded that specific 3D photoionization modelling of A\,30 is most needed in order to understand these differences.

To our knowledge, \citet{Dinerstein1984} were the first to suggest the existence of a disk-like structure surrounding the CSPN of A\,30. 
This structure, unveiled by near-IR observations, was suggested to be tracing knots J2 and J4 as later found by \citet[][]{Borkowski1994}. 
Subsequently, narrow-band images of the H-deficient clumps mainly, obtained through the [O\,{\sc iii}] 5007~\AA\, filter, have been used to described them as part of a disrupted disk and a couple of bipolar ejections \citep{Borkowski1993,Harrington1996}.

We note that there is no detailed morpho-kinematic study of A\,30 in the literature, although several kinematic studies indeed support the expanding disk and jet structure \citep[e.g.,][]{Reay1983,Jacoby1989,Chu1997,Yadoumaru1994}. 
The study presented by \citet{Meaburn1996} disclosed a more detailed kinematics given the large number of spectra and the higher spectral resolution. These authors used San Pedro M\'{a}rtir (SPM) Manchester Echelle Spectrograph \citep[MES;][]{Meaburn2003} observations to show that the radial expansion velocity of the outer H-rich nebula is 38.5$\pm$1.0 km~s$^{-1}$, while the H-poor material exhibits a range of velocities from $\sim$40 to a few times 100~km~s$^{-1}$.

The mass ejection from a single star can not produce a bipolar morphology \citep[e.g.,][]{GS2014}. Therefore, the bipolar structure of the H-deficient clumps in A\,30 (disrupted disk+jet) has to be attributed to the action of a companion, either stellar or, as suggested a few decades ago by \citet{Soker1997}, sub-stellar. Indeed A\,30 possesses the highest ADF among PNe, a property strongly correlated with binarity \citep{Wesson2018}, and the brightness variation period of 1.06~d of its CSPN in {\it Kepler/K2} observations has been attributed to the presence of a companion \citep{Jacoby2020}. Nevertheless, \citet{Toala2021} used photoionization models including gas and dust to find that, after accounting for the C trapped in the dust, the C/O ratio of $\gtrsim1$ is consistent with a single star experiencing a VLTP \citep[see, e.g.,][and references therein]{MB2006}.

In this paper, we present the first morpho-kinematic study of the inner H-deficient clumps in A\,30 to assess the details of their ejection scenario and shaping. High-resolution SPM MES spectra were recently acquired and modelled with the use of the {\sc shape} software \citep{Steffen2011}.

The paper is organized as follows. In Section~\ref{sec:obs} we present our observations and data reduction. The interpretation of the high-resolution spectra through a morpho-kinematic model is presented in Section~\ref{sec:shape}. Our results are discussed in Section~\ref{sec:diss} where we propose a shaping mechanisms of the H-deficient clumps. Finally, our conclusions are presented in Section~\ref{sec:conclusions}.

\section{Observations and data preparation}
\label{sec:obs}

Long-slit high-dispersion optical spectra of A\,30 were obtained using the MES mounted on the 2.1m telescope at the Observatorio Astron\'{o}mico Nacional at San Pedro M\'{a}rtir (OAN-SPM, Mexico)\footnote{The Observatorio Astronómico Nacional at the Sierra de San Pedro M\'{a}rtir (OAN-SPM) is operated by the Instituto de Astronom\'{i}a of the Universidad Nacional Aut\'{o}noma de M\'{e}xico.}. The data were obtained on 2020 January 20 and 21 (see Table~\ref{tab:obs}) using the E2V 42-40 CCD with a pixel size of 13.5 $\mu$m pix$^{-1}$ and a 4$\times$4 on chip binning, which resulted in a plate scale of 0.702 arcsec pix$^{-1}$. 
A series of spectra were taken with exposures of 1200 s through the [O\,{\sc iii}] 5007\AA\, filter ($\Delta\lambda$=50\,{\AA}) to isolate the 114th order (0.086\,{\AA} pix$^{-1}$ spectral scale). The slit width of 150 $\mu$m (1.9 arcsec) leads to a spectral resolution of $12\pm1$~km~s$^{-1}$.

\begin{table}
\setlength{\tabcolsep}{2.0\tabcolsep}
\caption{SPM MES observations of A\,30. 
All spectra have exposure times of 1200~s and were taken through a filter that isolates the [O\,{\sc iii}] $\lambda$5007 \AA\ emission line (see text for details).}
\label{Slits}
\centering
\label{tab:obs}
\begin{tabular}{cccc} 
\hline
Slit  &  PA           & Date          & seeing \\
      &  ($^{\circ}$) & (yyyy mmm dd) & ($''$) \\
\hline
1 & 60    & 2020 Jan 20 & 3.0 \\
2 & 40    & 2020 Jan 20 & 3.0 \\
3 & 20    & 2020 Jan 20 & 3.0 \\
4 & $-$30 & 2020 Jan 21 & 2.0 \\
5 & $-$45 & 2020 Jan 21 & 2.0 \\
\hline
\end{tabular}
\end{table}

\begin{figure*}
\begin{center}
\includegraphics[width=0.95\textwidth]{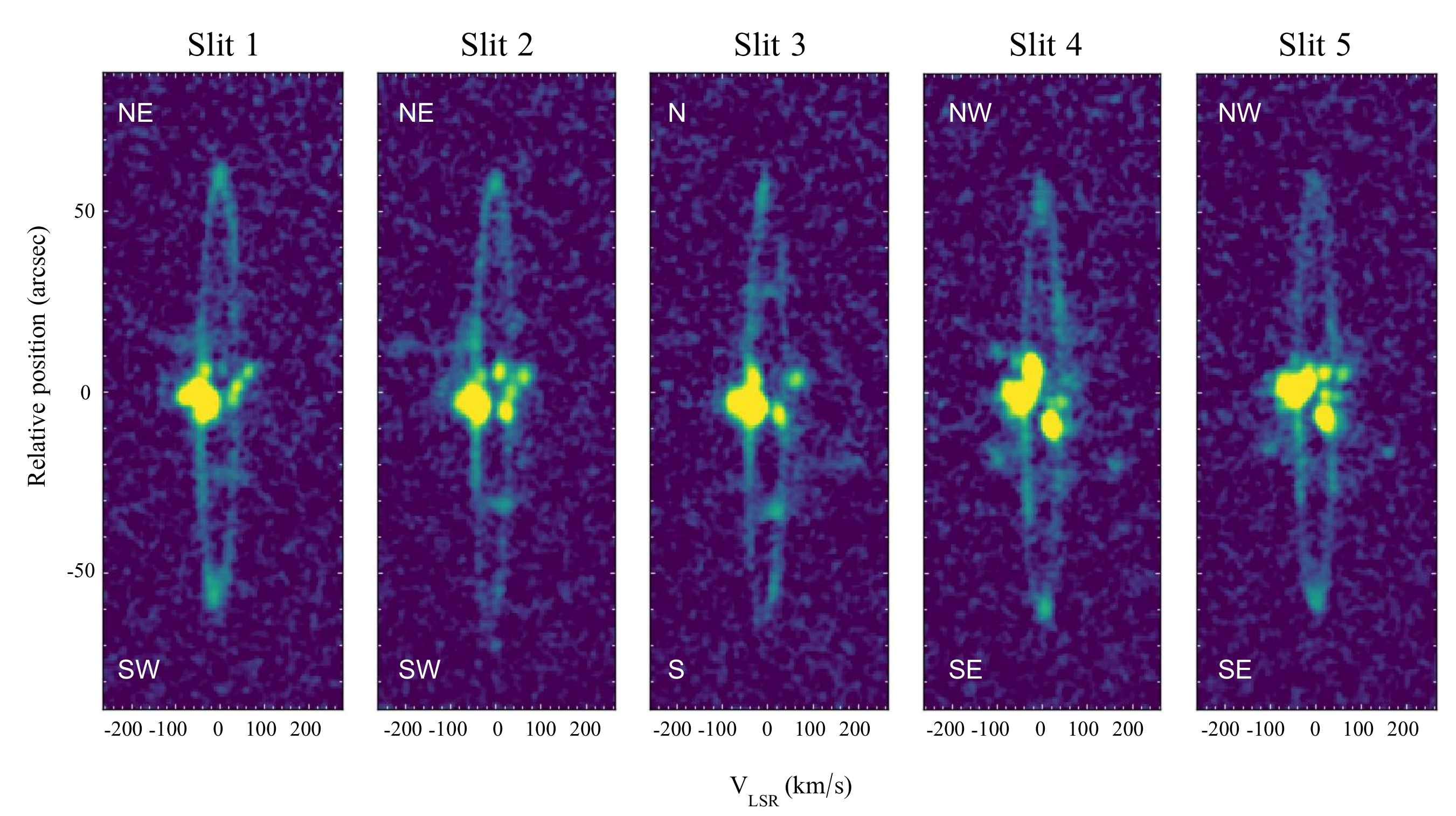}
\vspace*{-0.4cm}
\caption{PV diagrams obtained from the SPM MESS [O\,{\sc iii}] spectra of A\,30 for slits 1--5.}
\label{fig:pv_full}
\end{center}
\end{figure*}

\begin{figure*}
\begin{center}
\includegraphics[width=0.92\textwidth]{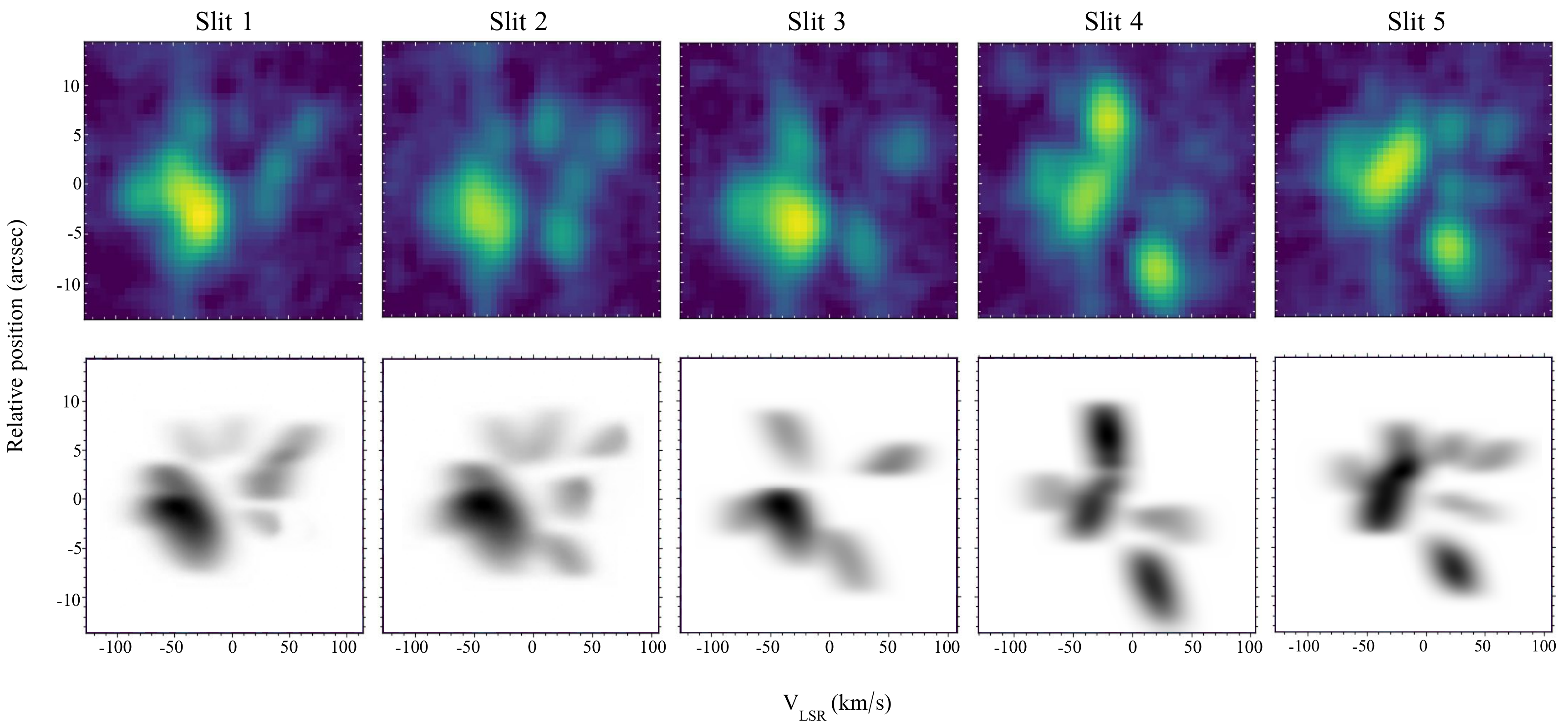}
\vspace*{-0.4cm}
\caption{Top panels: PV diagrams obtained from the SPM MES [O\,{\sc iii}] spectra of the central region of A\,30. Bottom panels: Synthetic PV diagrams obtained from our {\sc shape} model (see Sec.~\ref{sec:shape}).}
\label{fig:pv2}
\end{center}
\end{figure*}

We observed five slits positions across the CSPN of A\,30 along different position angles (PAs) to cover most of the H-deficient clumps and filaments surrounding the CSPN. 
The slit positions are illustrated in Fig.~\ref{fig:slits} and the journal of observations is given in Table~\ref{tab:obs}. 

The spectra were processed using standard calibration routines in {\sc iraf} \citep{Tody1993}, including bias subtraction and wavelength calibration with ThAr arc lamps obtained  immediately before and after the science observations. 
The wavelength accuracy is estimated to be $\pm$1 km~s$^{-1}$.

The resultant position-velocity (PV) diagrams\footnote{The PV diagrams were produced with the matplotlib {\sc python} routines developed by \citet{Hunter2007}.} of the five spectra are presented in Fig.~\ref{fig:pv_full}. These resemble PV diagrams previously presented by other authors, with an outer almost spherical shell extending $\approx$1 arcmin in radius and high-velocity structures ($\lesssim$200~km~s$^{-1}$) contained within the inner 30 arcsec (see Section~\ref{sec:intro}). To highlight the structures close to the CSPN of A\,30, we show close-ups of the PV diagrams in Figure~\ref{fig:pv2}-top.

\begin{figure*}
\begin{center}
\includegraphics[width=\textwidth]{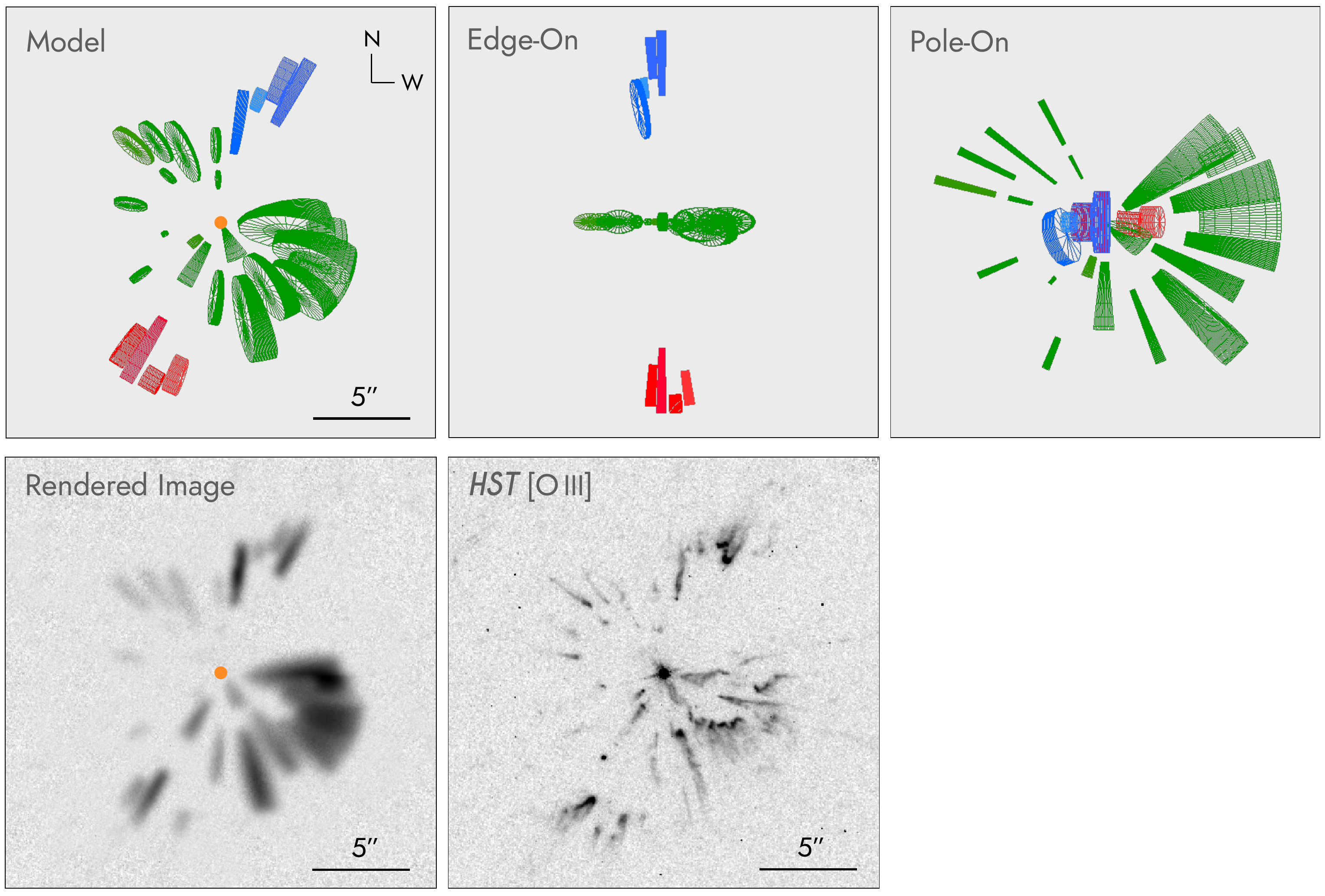}
%\vspace*{-0.4cm}
\caption{
Morpho-kinematic model of A\,30 obtained with the software {\sc shape}. 
Top panels show our best model through different viewing angles where features in green, blue, and red denote the disrupted disk and the approaching and receding jet-like structures in A\,30, respectively (see Section~\ref{sec:shape} for details). 
The bottom left panel shows a rendered image which is compared with the {\it HST} [O\,{\sc iii}] image in the bottom right panel. The position of the CSPN is illustrated with an orange dot in the leftmost panels. The edge- and pole-on images do not have the same scale as the other panels, they are presented for illustrative purposes only.}
\label{fig:shape}
\end{center}
\end{figure*}

\section{Morpho-kinematic model}
\label{sec:shape}

The software {\sc shape} was used to interpret the SPM MES [O\,{\sc iii}] spectra and {\it HST} WFC3 [O~{\sc iii}] image of A\,30.  
{\sc shape} allows the user to define different 3D geometric structures that can be edited (truncated, rotated and stretched) in order to produce shapes that resemble those of the direct images of the studied objects. 
Moreover, expansion velocity laws can be defined for the different structures and, subsequently, synthetic PV diagrams can be produced and compared to those observed.

A simplistic model consistent of a toroidal structure and a pair of spherical clumps mimicking the jet-like features was first attempted. 
Not surprisingly, those structures did not produce images nor PV diagrams similar to the observed ones as the H-deficient clumps and filaments in A\,30 are actually composed by collections of discrete knots. 
Furthermore, it was clear that the clumps and filaments in the so-called disrupted disk are not equidistant to the CSPN of A\,30. 
To improve the model, tailor-made structures were defined for each individual clump. 
These consisted of truncated tori, i.e., structures with the appearance of a doughnut slice.  
Those corresponding to the disrupted disk were assumed to be distributed mostly along the same plane, that will be referred as the equatorial plane, whereas those of the jet-like features were assumed to be mostly along the orthogonal direction, i.e., the symmetry axis of the disrupted disk.  
The model further assumes that the symmetry axis of the disrupted disk and thus the jet-like features have a PA on the sky of $-30^\circ$, the same used for slit 4 (see Table~\ref{tab:obs}).

The shape of each individual structure and their velocities were then fine-tuned until the features in the synthetic [O~{\sc iii}] image and PV diagrams were satisfactorily reproduced. 
Each truncated torus is defined by its distance to the star, thickness, length along the radial direction, and height over the equatorial plane. 
The velocity at each point of these truncated tori is assumed to be radial and its modulus $v$ is proportional to its radial distance $r$
\begin{equation}
    v(r) = v_0 \left(\frac{r}{r_0}\right),
\end{equation}
\noindent where $r_0$ is the distance from the facing-side of the clump to the CSPN and $v_0$ is a characteristic velocity of each clump.  
This velocity law thus includes a velocity gradient within each clump that takes into account the head to tail velocity difference in a cometary knot.

The best {\sc shape} model of A\,30 is presented in Figure~\ref{fig:shape} and the resultant synthetic PV diagrams are presented in the bottom panels of Figure~\ref{fig:pv2} in comparison with those observed. 
Edge- and pole-on projections of the best {\sc shape} model are also presented in the top-middle and top-right panels of Figure~\ref{fig:shape}.

This model includes 20 more or less coplanar truncated tori around the CSPN illustrated in green in Figure~\ref{fig:shape}.  
These structures are located on a plane with an average inclination angle $i$ of $37^\circ\pm4^\circ$ with respect to the line of sight, they have distances from the CSPN of A\,30 ranging between 2.1 and 11.4 arcsec, with an averaged distance of 8.3 arcsec, and the average disk expansion velocity has a value $V_\mathrm{disk}$ of $105\pm5$ km~s$^{-1}$, where the errors on $i$ and $V_\mathrm{disk}$ are derived empirically solely based on the accuracy in reproducing the observed PVs and image by the model. 
The model requires the brightest SW optical knots to be slightly larger (see Fig.~\ref{fig:shape}). 
Interestingly, the characteristic velocity of each clump $v_0$ decreases with their distance to the CSPN, in agreement with the results obtained by \citet{Fang2014} based on their proper motions. 
These structures fairly reproduce the morphology and kinematics of the disrupted disk\footnote{
We note that this comparison is limited by the lower spatial resolution of the kinematic data compared to the superb {\it HST} image quality.  
} as illustrated in the bottom panels of Figures~\ref{fig:pv2} and \ref{fig:shape}.

Meanwhile the jet-like features are reproduced using two groups of truncated tori, with the farthest ones along a direction orthogonal to that of the disrupted disk. 
These are located $\sim$24 arcsec from the CSPN and are illustrated in blue and red in Figure~\ref{fig:shape} for the approaching and receding structures, respectively. 
As the jets are assumed to be orthogonal to the disrupted disk, the best fit model implies a deprojected expansion velocity for the jet clumps $V_\mathrm{jet}$ of $60\pm5$ km~s$^{-1}$ for the SE one and $45\pm5$ km~s$^{-1}$ for the NW one.   
The largest expansion velocity of the SE jet is also in agreement with its largest proper motion \citep{Fang2014}.

\begin{figure}
\begin{center}
\includegraphics[width=\linewidth]{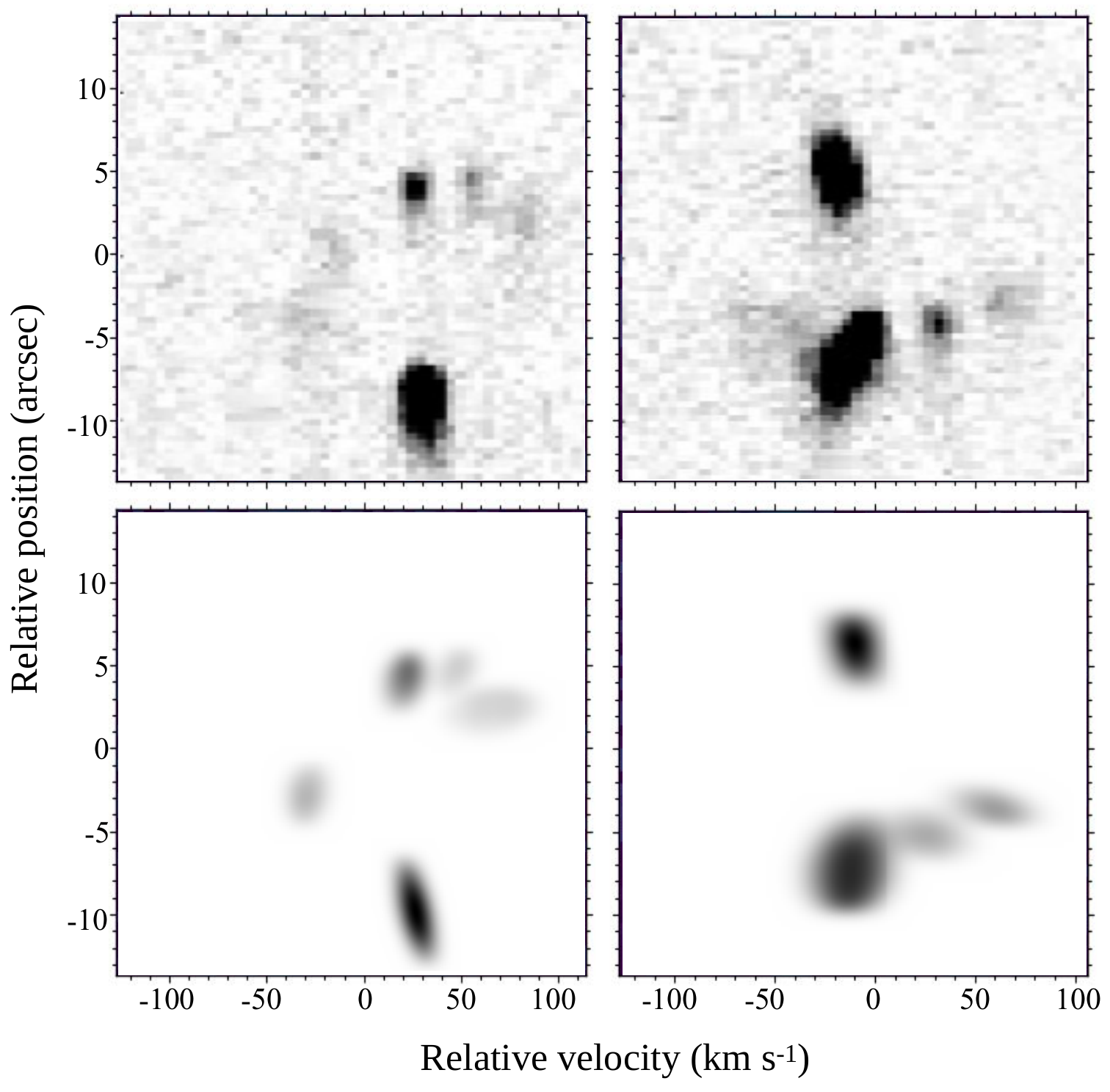}
\caption{Top panels: High-dispersion [O\,{\sc iii}] spectra of A\,30 presented in \citet{Chu1997} obtained at the 4 m Kitt
Peak National Observatory. Bottom panels: Synthetic spectra obtained from our best {\sc shape} model of A\,30. The spectra were obtained with a PA=0$^{\circ}$ in a N-S direction passing through the SE polar knot (left panels) and the NW polar knot (right panels).}
\label{fig:YHC}
\end{center}
\end{figure}

To further test our model, we also produced synthetic PV diagrams along long-slits similar to those presented by \citet{Chu1997}. These were 1~arcsec wide and were oriented N-S at offsets 4~arcsec East and West of the CSPN of A\,30, thus registering the SE and NW jet-like structures (knots J1 and J3), respectively, but also the eastern and western tips of the disrupted disk. The synthetic PV diagrams, shown in Figure~\ref{fig:YHC}, exhibit very similar spatio-kinematic patterns with those presented in figure 4 of \citet{Chu1997}.

\section{Discussion}
\label{sec:diss}

Within the model limitations, the morpho-kinematic {\sc shape} model described above does an excellent job reproducing the observed PV diagrams and image of A\,30 (see Figs.~\ref{fig:pv2} and \ref{fig:shape}). 
It definitely confirms that the H-deficient clumps surrounding the CSPN of A\,30 are distributed on a disrupted disk and a pair of jet-like features.

Contrary to what would be expected from the kinematics of a disk and a bipolar jet produced at the same time during a thermonuclear explosion, our {\sc shape} model indicates that the clumps in the disk-like structure in A\,30 expand faster than the jet features, while they are three times closer to the CSPN than those. 
These results are in line with those reported by \citet{Fang2014} who analyzed multi-epoch {\it HST} WFPC2 and WFC3 images of A\,30 and A\,78 and showed that the proper motions of the H-deficient clumps in these two born-again PNe are higher for those closer to the CSPN. A number of different physical mechanisms such as photoionization, photoevaporation, the rocket effect and momentum transfer from the current fast wind to the dense clumps may have contributed altering the dynamics of the H-deficient ejecta and producing its complex morphology \citep[see the discussion section in][]{Fang2014}.

The extreme bipolar morphology of the H-deficient clumps around the CSPN of A\,30 seems to concur with the possible presence of a companion \citep{Soker1997,Jacoby2020}, which can be invoked to explain the modulation of the mass loss and the production of the disk+jet system. 
Nevertheless, the C/O ratio obtained after taking into account the C-rich dust in the ejecta in A\,30 suggests a single-star VLTP event \citep{Toala2021}. 
We propose that these apparently discrepant results can be simultaneously explained by invoking a VLTP followed by a common envelope (CE) evolution \citep{Ivanova2013}.
This scenario, illustrated in Figure~\ref{fig:a30_scheme}, is detailed as follows:

\begin{figure}
\begin{center}
\includegraphics[width=0.9\linewidth]{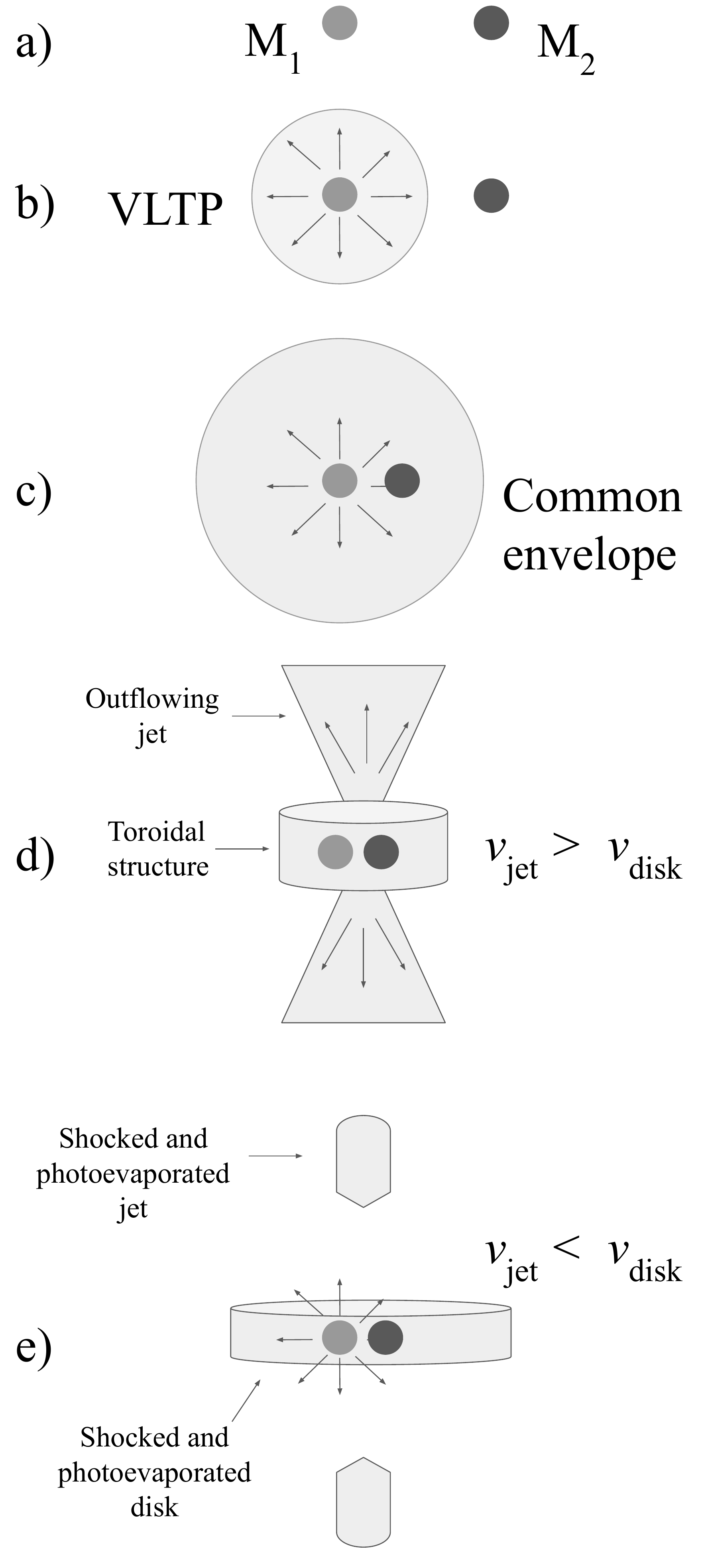}
\vspace*{-0.4cm}
\caption{Cartoon of the evolutionary path of the H-deficient ejecta in A\,30. This figure describes the CEBA scenario proposed in this work, in which a WD experiences a VLTP and the ejection of material is shaped into a bipolar morphology by a companion entering a CE phase. After the VLTP, the new WD develops a strong wind and ionizing photon flux that imprints momentum onto the H-deficient material. The closest structure is then accelerated faster than the bipolar structures. See Sec.~\ref{sec:diss} for details.}
\label{fig:a30_scheme}
\end{center}
\end{figure}

\begin{enumerate}[a)]%[label=(\alph*)]%[\hspace*{0.5cm}]

\item 
The CSPN of A\,30 evolved from a binary system comprising an evolved WD (white dwarf)
%post-AGB star 
and a low-mass companion with masses $M_1$ and $M_2$, respectively (Fig.~\ref{fig:a30_scheme}-{\it a}).

\item 
Following stellar evolution models, the WD experienced a VLTP causing the expansion of its outer layers and returning to the AGB locus of the HR diagram \citep[see, e.g.,][Fig.~\ref{fig:a30_scheme}-{\it b}]{Iben1983}. 
These predict a duration for the VLTP of decades to a few hundreds of years \citep[see, e.g.,][and references therein]{MB2006}.

\item 
The radius increase of the primary caused the system to enter a CE phase (Fig.~\ref{fig:a30_scheme}-{\it c}). 
The companion transferred angular momentum to the envelope, producing a reduction of its gravitational binding energy.
CE numerical simulations predict that this process reduces the orbital separation within time scales of months-to-years \citep[see, e.g.,][]{Chamandy2020}. 
We refer to this stage as the CE-born-again (CEBA) phase.

\item 
Numerical simulations have shown that the companion ($M_2$) can help shaping the mass-loss from the primary ($M_1$) during the CE phase. 
This process might have produced a toroidal structure followed by a subsequent bipolar ejection of material \citep[see, e.g.,][and references therein]{GS2021,Zou2020,LC2021}. 
Thus, in A\,30 the bipolar shaping took place during the duration of the VLTP.
At this moment, the jet velocity $V_\mathrm{jet}$ was larger than the disk expansion velocity $V_\mathrm{disk}$ (Fig.~\ref{fig:a30_scheme}-{\it d}). As the ejecta in the bipolar component moved faster than the disk, it was located at larger distances from the CSPN than the toroidal structure. This phase will be referred as early born-again (EBA) phase.

\item 
A large amount of works on born-again PNe suggests that the evolution of their CSPN is fast after experiencing a VLTP \citep[see][and references therein]{Evans2020,Hinkle2020}. The star shrinks and gets hotter so that a fast stellar wind and strong UV flux are developed.
We propose that after the EBA phase, the increased stellar wind momentum and ionizing flux induced the toroidal structure to experience hydrodynamical instabilities that ultimately disrupted it into clumps and filaments within a few hundred years \citep[see the simulations presented in][]{Fang2014}. 
The momentum transfer efficiency from the wind to the H-deficient material would be larger for closer structures. 
In particular, the CSPN of A\,30, which experienced a VLTP about $\lesssim$900~yr ago, has a current stellar wind with a terminal velocity of $\approx$4000~km~s$^{-1}$ and an effective temperature of 115~kK \citep{Guerrero2012}. 
Accordingly, the clumps in the disk that were closer to the CSPN were accelerated faster than the bipolar structures, ultimately producing the current kinematic configuration where the jet velocity $V_\mathrm{jet}$ is smaller than the disk expansion velocity $V_\mathrm{disk}$ (Fig.~\ref{fig:a30_scheme}-{\it e}). We will refer to this stage as late born-again (LBA) phase and propose that A\,30 can be actually placed into this evolutionary stage.

\end{enumerate}

The proposed scenario can also help explaining the fact that the larger and brighter clumps and filaments in the disrupted disk of A\,30 have a preferential distribution towards the SW of the CSPN. 
Simulations have shown that the toroidal structure formed as the result of a CE evolution is uneven and inhomogeneous \citep[see, for example, figure 1 in][and their associated videos]{Ondratschek2021}. Density enhancements appear and get diluted towards the equator of the toroidal structure in short time-scales and are not symmetric given the orbital motion of the companion star. Thus, it is very likely that by the time the fast wind and the UV flux of the CSPN of A\,30 increased, a density enhancement existed towards the SW.

In the proposed CEBA evolution scenario (see Fig.~\ref{fig:a30_scheme}), the disk and jet abundances would be the same and thus we suggest that the differences in their ADFs reported by \citet{Simpson2022} should be rather attributed to their different physical properties. These would produce variations in the extent of the temperature \citep{Peimbert1967} and density fluctuations \citep{Viegas1994}, or in the clumping of the H-deficient material \citep{Liu2000}, which have long been suggested to cause the ADF \citep[see, e.g.,][and references therein]{Wesson2018}. The IR study of A\,30 presented by \citet{Toala2021} indeed 
points to different physical properties between the disk and jet, with dust at temperatures of $\lesssim$200~K mainly present in the disrupted disk of A\,30. Alternatively, the interaction of the companion with the expanded star envelope during the CEBA and EBA phases, which would feed the disk with processed material, may result in chemical abundances different for the jet and disk. 
As concluded by \citet{Simpson2022}, it is critical to assess the specific mass fraction of H-poor and H-rich material of the different components of the ejecta to peer into this problem.

A detailed 2D mapping of the chemical abundances for each clump in A\,30 would resolve this issue, and appropriate high-dispersion integral field spectroscopic data of A\,30 are been currently sought.  
In addition, radiation-hydrodynamic numerical simulations following in detail the stellar evolution of stars experiencing a VLTP will be used in the future to test our proposed CEBA scenario and to assess the time scales on the formation of the complex structures in born-again PNe (Rodr\'{i}guez-Gonz\'{a}lez et al. in prep.).

\subsection{Comparison with other born-again PNe}

The proposed CEBA scenario can be used to describe the characteristics of other born-again PNe. 
In particular, we note that A\,78 shares many similarities with A\,30. The morphology of the dense H-deficient clumps in A\,78 also resemble a disrupted disk+jet system and their expansion parallaxes and radial velocity patterns are very similar to those of A\,30 \citep[e.g.,][]{Meaburn1998}, 
as well as their kinematic ages, $\lesssim10^{3}$~yr \citep{Fang2014}. 
In addition, their CSPNe exhibit almost the same properties \citep[see][]{Toala2015}. 
Thus, it can be envisaged that the CEBA scenario proposed for A\,30 would also explain the properties of A\,78.  
A detailed morpho-kinematic model of A\,78 is needed to fully confirm these suggestions.

The born-again PN A\,58, which experienced its VLTP in 1919, about 100~yr ago, also displays a bipolar ejection with a dense toroidal structure almost obscuring its CSPN \citep{Hinkle2008}. 
The expansion velocity of the optical ejecta has been shown to be $\sim$200~km~s$^{-1}$ \citep{Clayton2013}, but recent ALMA observations of the molecular component of A\,58 demonstrated that the toroidal structure expands with a velocity of 90~km~s$^{-1}$ from the CSPN whilst the bipolar emission has a faster velocity of 280~km~s$^{-1}$ \citep{Tafoya2022}. 
Interestingly, these authors found that the molecular material in the bipolar ejection can be traced closely to the CSPN and has a kinematic age of $\lesssim$20~yr, i.e., apparently younger than the VLTP event. 
The effects of the fast stellar wind and strong UV flux from the CSPN of A\,58 are not yet dominant as in A\,30 and A\,78. 
Thus, under our proposed scenario, A\,58 can be placed in the EBA phase of the evolution of a VLTP ejecta (Fig.~\ref{fig:a30_scheme}-{\it d}).

The youngest born-again PN, the Sakurai's Object, experienced its VLTP event in February 1996, about 26~yr ago.  
Only two years later, in March 1998, material expanding at velocities up to $-550$ km~s$^{-1}$ was revealed by a blue-shifted absorption line of He~{\sc i} 10830 \citep{Eyres1999}.  
Soon afterwards, optical observations obtained on 2001 detected a bipolar outflow in the [N~{\sc ii}] emission lines with two components at systemic velocities of $-350\pm50$ and $+200\pm50$ km~s$^{-1}$ \citep{Kerber2002}, thus implying an average velocity close to the 290 km~s$^{-1}$ derived from the fundamental vibration–rotation CO band lines around 4.7 $\mu$m detected by \citet{Eyres2004} in 2003.  
This expansion velocity seems to have remained unchanged in the last decade \citep{HJ2014,Hinkle2020}. 
Subsequent modelling of the IR emission led \citet{Chesneau2009} to suggest the presence of a disk-like structure seen almost edge on, obscuring the progenitor star.  
Interestingly, the expansion velocity of the densest molecular material probed by the sub-mm detection of the HCN molecule is $\gtrsim$100 km~s$^{-1}$ \citep{Tafoya2017}.  
If we assume that the densest molecular material is associated with the toroidal structure, its expansion velocity would be several times smaller than that of the bipolar outflow.  
This would place Sakurai's Object in the EBA phase, where the bipolar ejection is still faster than the disk-like structure. 
Our team has gathered ALMA data to map the molecular and continuum emission of Sakurai's Object to probe the extreme bipolar morphology of this born-again PN (Tafoya et al. in prep.).

The last addition to the born-again PN class is HuBi\,1 \citep{Guerrero2018,Montoro-Molina2022}. The 2D spatio-kinematic properties of its H-poor ejecta have been interpreted as a prolate ellipsoidal shell \citep{Rechy-Garcia2020}, although the extremely bright emission around the systemic velocity and weakness of the fastest components suggests an alternative physical model consistent of a bright equatorial disk mostly on the plane of the sky and a bipolar outflow tilted with the line of sight.  
The similarity between the expansion velocity of the disk on the plane of the sky\footnote{
We note that this value depends on the rather uncertain distance to HuBi\,1.
}
of 290 km~s$^{-1}$ and that of the bipolar outflows of 250 km~s$^{-1}$  \citep{Rechy-Garcia2020} suggests that HuBi\,1 would be in the transition between the EBA and LBA phases. Indeed HuBi\,1 experienced its VLTP about 200 yrs ago \citep{Rechy-Garcia2020} and thus it has an intermediate age between A\,58 and the more evolved A\,30 and A\,78.

\subsection{Model caveats}

The proposed morpho-kinematic model and CEBA scenario provide a reasonable description of the morphology and spatio-kinematics of A\,30 and its formation and evolution.  
Still we would like to mention a few caveats.

The availability of a throughout investigation of the proper motion of the knots of A\,30 using multi-epoch {\it HST} WFPC2 and WFC3 images \citep{Fang2014} allows investigating their tangential velocity and thus their inclination in conjunction with the observed radial velocities.  
Unfortunately, the knots of the disrupted disk are not well resolved in our morpho-kinematic data and they cannot be assigned to the corresponding knot resolved in the {\it HST} images.  
This investigation is still viable for the polar knots. 
The proper motions of NW (J3) and SE (J1) clumps is reported to be 7.93 and 6.99 mas~yr$^{-1}$, respectively.  
At the updated distance to A\,30 of 2.08$\pm$0.14 kpc \citep{BailerJones2021}, this implies tangential velocities of 69 km~s$^{-1}$ for the NW jet (J3) and 78 km~s$^{-1}$ for the SE jet (J1). 
The observed averaged systemic radial velocities of these jets are $-15\pm5$ km~s$^{-1}$ and $+21\pm5$ km~s$^{-1}$, respectively, where the error is introduced to account for their velocity gradient in the PV maps. 
The tangential and radial velocities imply expansion velocities and inclinations with the plane of the sky of $71 \pm 2$ km~s$^{-1}$ and 12$^\circ\pm$4$^\circ$ for the NW jet, and $81 \pm 2$ km~s$^{-1}$ and 15$^\circ\pm$4$^\circ$ for the SE jet. 
These results indicate that the SE jet moves faster than the NW jet, as in the morpho-kinematic model presented Section~3, but, unlike in the model, the polar outflows are not orthogonal to the disrupted disk, which would require an inclination of $37^\circ$. 
This is not a major issue, as the inclination of the polar outflow in the model is an independent parameter that can be fine-tuned together with the expansion velocity of the outflow. 
Interestingly, the bipolar ejection in Sakurai's Object is neither orthogonal to its equatorial disk \citep[see figure 2 in][]{Hinkle2020}. 
We note that the field of study of jet evolution and effects during the CE phase is very active, with simulations showing deviations from a linear motion and choking effects in jets \citep[see, e.g.,][and references therein]{Zou2022} that are not taken into account in the proposed CEBA scenario.

The comparison of the different born-again PNe in the previous section has evidenced that the H-poor ejecta in all of them can be interpreted as the combination of a disk and a jet.  
Their expansion velocities, however, are very different from one born-again PN to another. 
The disks and jets of A\,30 and A\,78 exhibit much lower expansion velocities than those of the Sakurai's Object, A\,58 and HuBi\,1, event though the current fast stellar wind from their CSPNe are imprinting momentum to the H-deficient clumps. 
It shall be noticed that, in the CEBA scenario, the bipolar ejection is produced by the presence of a companion.  
At early times, the velocity of the bipolar ejection is more or less the escape velocity
\begin{equation}
    v_\mathrm{esc} = \sqrt{2 G \frac{M}{R}}, 
\end{equation}
associated to the companion producing the jet, whereas the expansion velocity of the disk would depend both on the mass of the companion and its final orbital radius.  
There are no estimations of the mass and/or spectral types of the companions of the CSPN of born-again PNe, neither of their orbital radii, but it can be easily envisaged that a variety of those can produce very different initial values of the disk and jet velocities in born-again PNe that would explain the large difference of observed expansion velocities. 
On the other hand, the released energy of the VLTP might not be the same for all born-again PNe, but we note that this idea seems more difficult to assess without modern stellar evolutionary models.

\section{Conclusions}
\label{sec:conclusions}

We presented the first morpho-kinematic model of the H-deficient ejecta in a born-again PN, namely A\,30. 
We obtained SPM MES long-slit high-dispersion observations in the emission line of [O~{\sc iii}] of the central H-deficient clumps that mapped the so-called disrupted disk and jets surrounding the CSPN of A\,30. 
The observations were interpreted in conjunction with an \emph{HST} WFC3 [O~{\sc iii}] image of A\,30 by means of the software {\sc shape}, which allowed us to produce a 3D model of the spatial distribution of the H-deficient ejecta and their expansion velocities.

Our best {\sc shape} model is able to explain the features unveiled by the [O~{\sc iii}] PV diagrams and direct image. 
To reproduce the spatio-kinematic signatures of the disrupted disk-like structure in A\,30, our model requires the presence of a collection of clumps located mostly on a plane tilted by $37^{\circ}\pm4^{\circ}$ with respect to the line of sight with averaged distance of 8.3~arcsec from the CSPN and a deprojected expansion velocity $V_\mathrm{disk}$ of $105\pm5$ km~s$^{-1}$. 
The jet-like features are modelled by a pair of groups of clumps aligned mostly orthogonally with the disrupted disk. 
These expand with velocities $V_\mathrm{jet}$ of $45\pm5$ km~s$^{-1}$  for the approaching NW component and $60\pm5$ km~s$^{-1}$ for the receding SE one. 
The combination of the tangential and radial velocities of these jet components provides expansion velocities and inclinations with the plane of the sky of $71 \pm 2$ km~s$^{-1}$ and 12$^\circ\pm$4$^\circ$ for the NW jet, and $81 \pm 2$ km~s$^{-1}$ and 15$^\circ\pm$4$^\circ$ for the SE jet. 
This would imply that the bipolar outflow is not orthogonal to the disk, but this assumption on the morpho-kinematic model can be relieved just adopting expansion velocities and inclinations similar to those reported above. 
Together with the reports on the larger proper motions of the clumps located closer to the CSPN \citep{Fang2014}, the results above suggest that the jets and the disrupted disk in A\,30 are apparently not coeval.  
Actually, they rather imply that their kinematics have been strongly affected by dynamical processes, which can be dominated by their photoevaporation and the transfer of momentum of the stellar wind.

We propose that the jets of A\,30 have been produced during a CE phase between its CSPN and a companion. 
When the CSPN of A\,30 experienced the VLTP, it might have inflated its outer layers causing a CE phase. 
The subsequent fast evolution of the CSPN of A\,30 after the born-again event produced the current high ionization photon flux and fast stellar wind, which affects more strongly to the structure close to the CSPN, i.e, the disk. 
We suggest that whereas the kinematic properties of the jet-like features are currently similar to those during the CE phase, the clumps in the disk have  experienced a strong acceleration caused by the fast stellar wind.

We propose that other born-again PNe can also be explained by the proposed CEBA scenario presented here. 
In particular, we suggest that A\,30 and A\,78 are at the same late stage of evolution, when wind momentum is efficiently transferred to the clumps in the disk and $v_\mathrm{jet} < v_\mathrm{disk}$. 
The younger born-again PNe A\,58 and Sakurai's Object are arguably at a previous stage, when their CSPNe have not developed a strong stellar wind nor a strong UV flux to imprint momentum into the H-deficient material so that $v_\mathrm{jet} > v_\mathrm{disk}$. 
HuBi\,1 would be in the transition between these two phases, with $v_\mathrm{jet} \simeq v_\mathrm{disk}$.

Future 3D radiation-hydrodynamic simulations following in detail the evolution of born-again scenarios in combination with a CE phase are needed to test the predictions presented here.
High-dispersion integral field spectroscopic observations of A\,30 and A\,78 are sought to produce a 2D map of the ADF and C/O abundances among the different components of the H-poor ejecta.  
The availability of the proper motions of the clumps in these two born-again PNe on the plane
of the sky \citep{Fang2014} provides the unique opportunity to determine the inclination with
the line of sight and true space velocity of each knot without any geometrical assumption to
produce the first completely unbiased 3D model of the ejecta of born-again PNe.

\section*{Acknowledgments} 

The authors thank the anonymous referee for comments and suggestions that help clarify the presentation of our results.
J.B.R.-G., E.S. and G.R. acknowledge support Consejo Nacional de Ciencia y Tecnolog\'{i}a (CONACyT, Mexico) for a student scholarship. J.A.T. acknowledges funding from the Marcos Moshinsky Foundation (Mexico) and Direcci\'{o}n General de Asuntos del Personal Acad\'{e}mico (DGAPA), Universidad Nacional Aut\'{o}noma de M\'{e}xico, through grants Programa 
de Apoyo a Proyectos de Investigaci\'{o}n e Inovaci\'{o}n Tecnol\'{o}gica (PAPIIT)
IA101622. J.A.T. also acknowledges support from the 'Center of Excellence Severo Ochoa' Visiting-Incoming program. J.A.T. and GR-L acknowledge support from CONACyT (grant 263373). 
M.A.G.R. and B.M.M. acknowledge support from the State Agency for Research of the Spanish MCIU
through the ‘Center of Excellence Severo Ochoa’ award to the
Instituto de Astrof\'{i}sica de Andaluc\'{i}a (SEV-2017-0709). 
Y.-H.C. acknowledges the support of MOST grant 110-2112-M-001-020 from the Ministry of Science and Technology of Taiwan. L.S. acknowledges support from UNAM DGAPA PAPIIT Grant IN110122 (Mexico).

This paper is based in part on ground-based observations from the Observatorio  Astron\'{o}mico Nacional at the Sierra de San Pedro M\'{a}rtir (OAN-SPM), which is a national facility operated  by the Instituto de Astronom\'{i}a of the Universidad Nacional Aut\'{o}noma de M\'{e}xico. The authors thank the telescope operator P.~F.\,Guill\'{e}n for valuable guidance during several observing runs, and to the OAN-SPM staff for their valuable support. This work has made extensive use of the NASA's Astrophysics Data System.

\section*{Data availability}

The data underlying this article will be shared on reasonable 
request to the corresponding author.

%%%%%%%%%%%%%%%%%%%%%%%%%%%%%%%%%%%%%%%%%%%%%%%%%%



%\bibliographystyle{mnras}


%%%%%%%%%%%%%%%%%%%%%%%%%%%%%%%%%%%%%%%%%%%%%%%%%%

%%%%%%%%%%%%%%%%% APPENDICES %%%%%%%%%%%%%%%%%%%%%

%\appendix



%%%%%%%%%%%%%%%%%%%%%%%%%%%%%%%%%%%%%%%%%%%%%%%%%%

% Don't change these lines
\bsp	% typesetting comment
\label{lastpage}
\end{document}